# Selective Functionalization of Halogens on Zigzag Graphene Nanoribbons: A Route to the Separation of Zigzag Graphene Nanoribbons

Hoonkyung Lee, Marvin L. Cohen, and Steven G. Louie[*]

Department of Physics, University of California at Berkeley, Berkeley, California 94720, USA

Materials Sciences Division, Lawrence Berkeley National Laboratory, Berkeley, California 94720, USA

*Corresponding author: sglouie@berkeley.edu.

## ABSTRACT

Using the ab initio pseudopotential density functional method, we investigate the functionalization of halogen molecules into graphene-based nanostructures with zigzag and armchair edges. We find that halogen molecules adsorb through chemisorption on the zigzag edge carbon atoms with a binding energy of ~1−5 eV and their adsorption on a perfect zigzag edge is preferred, in sharp contrast to physisorption on the armchair edge and elsewhere where they adsorb with a binding energy of ~0.07 eV. We suggest that our findings would be utilized for an approach to the separation of zigzag graphene nanoribbons with regular edges with the change of the solubility of the functionalized nanoribbons.

KEYWORDS. Zigzag graphene nanoribbons, halogen molecules, and selective functionalizations.

Graphene nanoribbons (GNRs) with less than 10 nm width, which form a one-dimensional honeycomb sp[2]-structured single layer of carbon atoms, have attracted much attention because of their intriguing electronic and spin transport properties.[1-5] Recently, GNRs have been synthesized experimentally using several methods such as unzipping carbon nanotubes (CNTs), lithographic patterning, and chemical narrowing[6-8], and n-type or p-type field effect transistors (FETs) based on GNRs have been made[9]. In addition, progress on making clean GNRs has been made[10]. This experimental progress shows the feasibility of a new type of nanoelectronics based on graphene. Even prior to the experimental investigations, theoretical studies have shown that armchair graphene nanoribbons (AGNRs) and zigzag graphene nanoribbons (ZGNRs) can be utilized as nanoelectric and nanophotonic devices because of their intrinsic bandgaps of ~0.5–3.0 eV.[11-15] The bandgap for AGNRs arises from quantum confinement effects[12] and the bandgap for ZGNRs is due to the antiferromagnetic interaction between magnetic edge states on the opposite edges[11,12]. Another striking feature is that, unlike AGNRs, ZGNRs with less than 10 nm can be half-metallic through time reversal symmetry breaking by an electric field.[16] So, ZGNRs can be applicable for electronics as well as spintronics. However, highly-irregular-edged ZGNRs are not expected to possess the half-metallicity property, which means that ZGNRs with relative straight edges are necessary for applications of spintronics based on graphene. Furthermore, it has been recently found that ZGNRs with straight edges decorated with calcium atoms can be employed for high-capacity hydrogen storage materials.[17] Since GNRs with variously-shaped edges are made concurrently by methods such as unzipping CNTs, the separation of ZGNRs with straight edges is an important task for practical applications as in the case of the separation of metallic and semiconducting CNTs.[18]

We note however that ZGNRs and AGNRs have different edge shapes whereas generally CNTs have the same surface shapes regardless of chirality. Therefore we suggest that some molecules may prefer to be functionalized on the differently shaped edges. In this paper, we investigate the functionalization of halogen molecules on graphene-based nanostructures with zigzag and armchair edges. We find that halogen molecules adsorb through chemisorption on the zigzag edge carbon atoms with a binding energy of ~1−5 eV, in sharp contrast to physisorption on the armchair edge or elsewhere where they adsorb with a binding energy of ~0.07 eV. These functionalizations could induce a difference in the solubility between functionalized and unfunctionalized GNRs, which could be utilized as an approach to the separation of ZGNRs.

Our calculations were carried out using the ab initio pseudopotential density functional method within the plane-wave-based total energy minimization scheme[19]. The exchange correlation energy functional in the generalized gradient approximation (GGA)[20] was used, and the kinetic energy cutoff was taken to be 30 Ry. The optimized atomic positions were obtained by relaxation until the Hellmann-Feynman force on each atom was less than 0.01 eV/Å. Supercell[21] calculations were employed throughout where the atoms on adjacent nanostructures were separated by over 10 Å to eliminate unphysical interactions between repeated structures in the periodic calculations.

To investigate the adsorption characteristics of the halogen molecules, fluorine ($F_2$), chlorine ($Cl_2$), bromine ($Br_2$), and iodine ($I_2$) on ZGNRs and AGNRs, we perform calculation of these halogen molecules on the edge or on the middle of a ZGNR with a width of 13.5 Å and an AGNR with a width of 17.8 Å. Figure 1a,b shows the atomic structures for halogen molecules adsorbed on the edge of the ZGNR and AGNR. (In all calculations, the dangling σ bonds of the edge atoms are passivated by hydrogen atoms.) We find that the halogen molecules adsorb on the edge carbon atoms of the ZGNR in the form of dissociated atoms with an adsorption geometry that is like $sp^3$-hybridization (Figure 1a). In contrast, they adsorb on the middle of the ZGNRs and the edge or the middle of the AGNRs in the form of nondissociated diatomic molecules (Figure 1b) except for the case of the adsorption of a $F_2$ molecule. This is consistent with a recent result[22] that $Br_2$ molecules dissociatively adsorb on the graphitic zigzag edge. In contrast, a $F_2$ molecule adsorbs not only on the edge of the ZGNRs and AGNRs but also on the middle of the ZGNRs and AGNRs in the form of dissociated atoms. However, the adsorption of a $F_2$ molecule on the edge of a ZGNR is preferred by ~3 eV compared to that on the middle of the ZGNRs or on the AGNRs. The large binding energy of halogen molecules in a dissociated atomic form on the zigzag edges is due to the zigzag edge's property associated with a low energy π-electron edge state[24], making it chemically more reactive than an armchair edge. The armchair edge energy is lower than zigzag edge energy by 0.2 eV/Å.[23] We also have confirmed that the adsorption of halogen molecules does not depend on which of the three families the AGNR is in. The calculated binding energy of halogen molecules for all the cases considered is presented in Table I. When $F_2$, $Cl_2$, $Br_2$, and $I_2$ molecules adsorb on the edge of ZGNRs after dissociation, the binding energy of the $F_2$, $Cl_2$, $Br_2$, and $I_2$ molecules has the order $F_2>Cl_2>Br_2>I_2$, and the distance between the F, Cl, Br, and I atom and the C atom is 1.8, 2.0, 2.2, and 2.4 Å, respectively. The distance between the halogen molecules and the nearest C atom is ~3.2 Å when halogen molecules adsorb on the middle of ZGNRs and on AGNRs without dissociation except for the case of the $F_2$ molecule, which is close to the equilibrium distance via the van der Waals interaction.

We also consider different zigzag edge geometries of the following graphene-based nanostructures: an armchair-zigzag-edged GNR, a large zigzag-edged vacancy-defected graphene, a ZGNR with an irregular edge, and a hydroxyl-functionalized ZGNR to examine the tendency for the adsorption of halogen molecules for different edge geometries. As in the case of ZGNRs, halogen molecules adsorb dissociatively on the local zigzag-like edge and the hydroxyl-functionalized zigzag edge as shown in Figure 1c−f. However, as seen in the case of $I_2$, the binding energy (0.06 eV/$I_2$: molecular adsorption) of the molecule on the local zigzag edges is significantly reduced compared to the value of 0.5 eV/$I_2$ on the edge of a ZGNR as presented in Table I. In contrast, the binding energy of the halogen molecules on the hydroxyl-functionalized zigzag edge is roughly the same as that on the edge of ZGNRs as presented in Table I. We also confirmed that the tendency for adsorption of halogen molecules on a local armchair edge and a functionalized armchair edge is the same as that for AGNRs. These results show that the tendency for the adsorption of halogen molecules on local zigzag and armchair edges is the same as that on the edges of ZGNRs and AGNRs, and, regardless of a $sp^2$-functionalization to the edges, only adsorption of halogen molecules on the long zigzag edge (i.e., ZGNR's edge) is preferred to elsewhere on graphene-based nanostructures. Another attractive feature is that $I_2$ molecules dissociatively adsorb only on the perfect zigzag edge, in contrast to the molecular adsorption on the other edges.

To investigate the interaction between the halogen atoms attached on the edge of ZGNRs, we have calculated the total energy for the adsorption of two halogen atoms on the edge of a ZGNR as the distance between adsorbed halogen atoms increases from 2 to 15 Å as shown in the inset of Figure 2a. The interaction between two halogen atoms is repulsive and is inversely proportional to the square of the distance between the halogen atoms as shown in Figure 2a. Regardless of the kind of molecules, the interaction is negligible when the two halogen atoms are separated by more than ~12 Å (Figure 2a). In addition, the configuration consisting of one halogen atom adsorbed on each side of the same edge with the halogen-halogen distance less than ~10 Å is lower by ~0.2 eV (per 2 halogen atoms) than the configuration when both halogen atoms adsorb on the same side of the edge. We also investigate the dependence of the binding energy on the width of the ZGNRs for the case when the atoms adsorb on alternating sides of the ZGNRs. The calculated binding energy of a halogen molecule is saturated as the width reaches ~18 Å as shown in Figure 2b. These results imply that the minimum width and length with negligible interaction of the halogen atoms is ~18 Å and ~12 Å, respectively.

To describe the equilibrium conditions between the halogen gas reservoir and the adsorbed halogen molecules, we consider the grand partition function which is given by $Z_i = \sum_{n=0}^{N_{max}} g_n^i e^{n(\mu_i - \varepsilon_n^i)/kT}$ where the maximum number of adsorbed molecules per the unit cell is $N_{max}$, $n$ is the number of adsorbed halogen molecules, $g_n$ is the degeneracy of the configuration for a given $n$, $\mu_i$ is the chemical potential of halogen gas where $i$ indicates the kinds of the gases, $-\varepsilon_n^i$ (>0) is the binding energy of halogen molecules, and $k$ and $T$ are the Boltzmann constant and the temperature, respectively. The (fractional) occupation number per unit cell is obtained from the relation of $f_i = kT \partial \log Z_i / \partial \mu_i$ [25]:

$$f_i = \frac{\sum_{n=0}^{N_{max}} g_n^i n e^{n(\mu_i - \varepsilon_n^i)/kT}}{\sum_{n=0}^{N_{max}} g_n^i e^{n(\mu_i - \varepsilon_n^i)/kT}} \quad (1)$$

To estimate the coverage density of halogen molecules on the edge of ZGNRs, we explore the thermodynamics of adsorption of halogen molecules on ZGNRs. We consider the adsorption of halogen molecules on the edge of a ZGNR with a width of ~18 Å and a repeated cell length of ~12 Å (which is the minimum width and length with no interaction for absorption of one halogen molecule) as a function of the number of adsorbed halogen molecules, where up to six halogen molecules per unit cell can be adsorbed as shown in Figure 3a. The calculated binding energy of the halogen molecules is presented in Table II, which is reduced as the number of adsorbed halogen molecules increases because of the repulsive interaction between the halogen atoms. Using Eq. (1) and the calculated energy ($-\varepsilon_n^i$), we calculate the occupation number of halogen molecules on the ZGNR where the chemical potential of the ideal gas for halogen gases was used. The calculated occupation number of the $F_2$, $Cl_2$, $Br_2$, and $I_2$ molecules on the ZGNR per unit cell at 300 K and 0.1 atm is 6.00, 2.00, 2.00, and 0.13, respectively. This is attributed to the Gibbs factor ($e^{n(\mu_i-\varepsilon_n^i)/kT}$) for the adsorption of the $F_2$, $Cl_2$, $Br_2$, and $I_2$ molecules for $n=6$, $n=2$, $n=2$, and $n=0$ respectively, which dominates at 300 K and 0.1 atm where the values of the chemical potentials for $F_2$, $Cl_2$, $Br_2$, and $I_2$ gas are –0.48, –0.50, –0.53, and –0.55 eV, respectively. At 150 K and 0.1 atm, where the value of the chemical potential of $I_2$ gas is –0.25 eV, the occupation number of $I_2$ molecules reaches up to 1.98. This implies that the coverage of $F_2$, $Cl_2$, $Br_2$, and $I_2$ on the edge of ZGNRs can reach 100, 33, 33, and 33%, respectively. Another attractive feature is that $Br_2$ and $I_2$ molecules adsorb only on the relatively perfect zigzag edge with 33% coverage at 300 K and 150 K under 0.1 atm, respectively, when variously edged GNRs are mixed because the Gibbs factor for the adsorption of $Br_2$ and $I_2$ molecules on the edge of ZGNRs dominates compared with that for the adsorption of $Br_2$ and $I_2$ on other edges (e.g., local zigzag edges or armchair edges) because their binding energy on the edge of ZGNRs is much greater than that on the other edges.

Now, we estimate the temperature and the pressure to release the halogen molecules adsorbed on ZGNRs. Figure 3b,c shows the occupation number of halogen molecules on the ZGNR as a function of the temperature and the pressure. Analysis of our results show that, basically, $F_2$ and $Cl_2$ molecules adsorbed on the ZGNR are desorbed at a temperature of more than ~700 K because the binding energy is too large, but, in contrast, $Br_2$ and $I_2$ molecules are desorbed by a temperature of ~550 and 250 K at ~$10^{-4}$ atm, respectively, which are readily achievable. This result shows that $Br_2$ and $I_2$ are suitable for the applications of separation of ZGNRs.

Our calculations demonstrate the selective functionalization of halogen molecules on the edge of ZGNRs. A recent experiment[25] has shown that fluorination takes place on CNTs and the solubility of the functionalized graphene and CNTs is changed in water. Fluorination changes the nonpolar CNTs to polar ones because of the charge transfer between the CNTs and the fluorine atoms, and this may be responsible for the change of the solubility of the functionalized CNTs in water.[26] The interaction between the zigzag edge and $F_2$, $Br_2$, or $I_2$ has been experimentally observed, resulting in diminishing the edge states.[24,27] Furthermore, it has been observed that fluorine prefers reaction with the carbon edges and all the edges are occupied by fluorine atoms, which is consistent with our results. These experiments indicate that the unusual selective functionalization of halogen molecules on the zigzag carbon edge we find is realistic and the separation of narrow ZGNRs by functionalization that could cause a solubility difference between functionalized ZGNRs and unfunctionalized GNRs in a solvent can be feasible as schematically shown in Figure 4. The coverage of the edge of ZGNRs by halogen atoms can be as much as 30−100%, which may be enough to change the solubility of the functionalized GNRs. Another striking feature is that the separation of *narrow* ZGNRs with straight edges may be possible because the adsorption of halogen molecules on the long zigzag edge is preferred and functionalized wider ZGNRs may not make the solubility difference large enough from the unfunctionalized GNRs because of their large areas of hexagonal carbon regime.

In conclusion, we have shown selective functionalization of halogen molecules on zigzag graphene nanoribbons compared to other ribbons using the ab initio pseudopotential density functional method. We propose that this functionalization could be utilized for an approach to separate narrow zigzag graphene nanorribons with straight edges from other nanoribbons by the induced solubility change. Bromine and iodine are suitable for this application because of their selective adsorption to zigzag graphene nanoribbons and readily achievable desorption temperature and pressure.

**ACKNOWLEDGMENT.** H. L. was supported by NSF Grant No. DMR07-05941. Numerical simulations were supported by the Director, Office of Science, Office of Basic Energy Sciences,

**Table I**. Calculated binding energy (eV/molecule) of a halogen molecule on the edge or the middle of a ZGNR, an AGNR, an armchair-zigzag-edged GNR (AZG), a large zigzag-edged vacancy-defected graphene (LZVG), a ZGNR with an irregular edge, and a hydroxyl-functionalized ZGNR (OH-ZGNR).

| Adsorption position | $F_2$ | $Cl_2$ | $Br_2$ | $I_2$ |
|---|---|---|---|---|
| **ZGNR-edge** | 5.30 | 1.73 | 1.17 | 0.50 |
| **ZGNR-middle** | 1.67 | 0.06 | 0.06 | 0.06 |
| **AGNR-edge** | 2.38 | 0.08 | 0.08 | 0.06 |
| **AGNR-middle** | 1.15 | 0.06 | 0.06 | 0.05 |
| **AZG-zigzag-edge** | 4.36 | 0.81 | 0.26 | 0.06 |
| **LZVG-edge** | 4.71 | 1.16 | 0.62 | 0.06 |
| **ZGNR-irregular-edge** | 4.40 | 0.89 | 0.40 | 0.06 |
| **OH-ZGNR-edge** | 5.42 | 1.49 | 1.00 | 0.39 |

**Table II**. Calculated binding energy (eV/molecule) of halogen molecules on the edge of a ZGNR as a function of the number of adsorbed halogen molecules. NA stands for not attachable.

| **Number of molecules** | $F_2$ | $Cl_2$ | $Br_2$ | $I_2$ |
|---|---|---|---|---|
| 1 | 5.30 | 1.73 | 1.17 | 0.50 |
| 2 | 5.17 | 1.60 | 1.06 | 0.40 |
| 3 | 4.67 | 1.13 | 0.61 | 0.12 |
| 4 | 4.07 | 0.63 | 0.18 | NA |
| 5 | 3.65 | 0.28 | NA | NA |
| 6 | 3.38 | 0.13 | NA | NA |

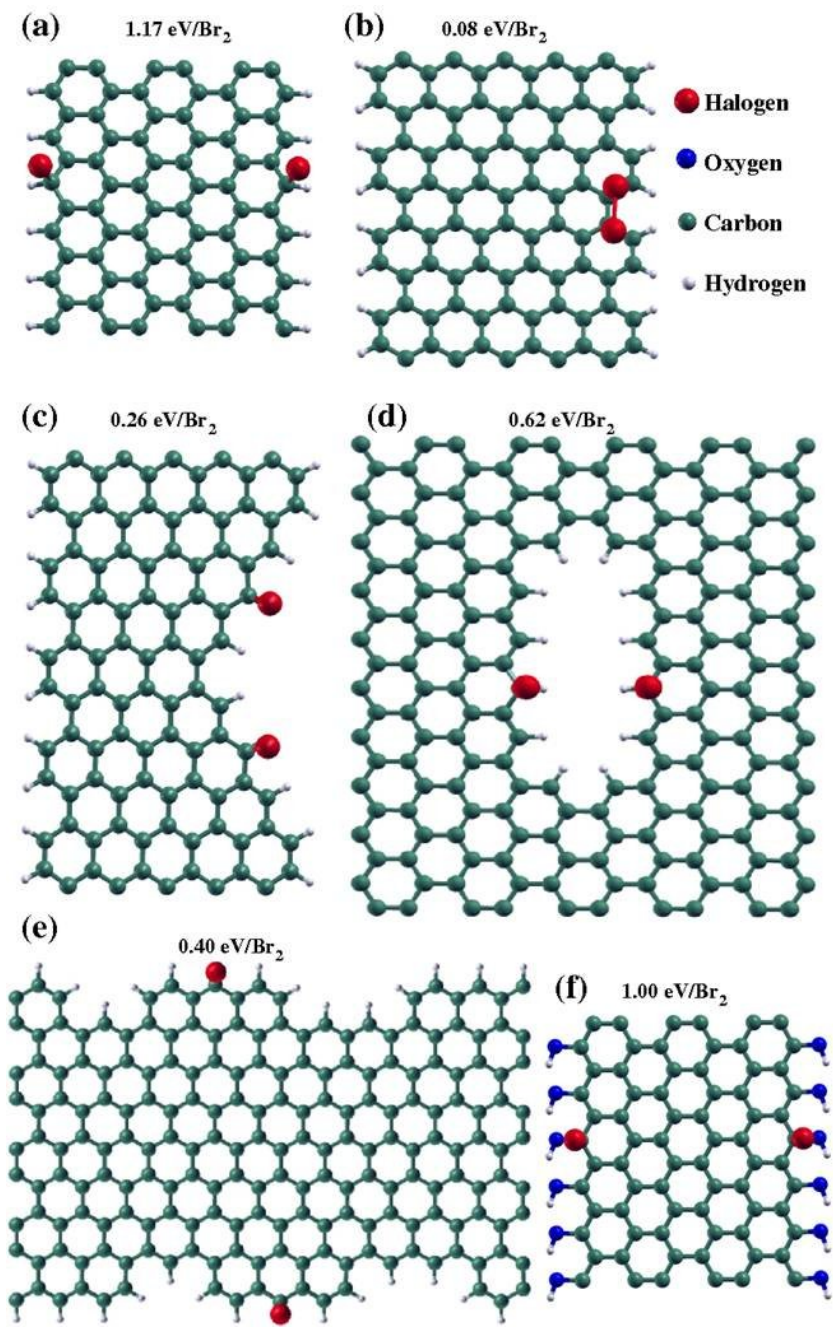

**Figure 1.** (a,b) The optimized atomic geometries for a halogen molecule adsorbed on the edge of a ZGNR and AGNR, respectively. (c-f) The optimized atomic structures for a halogen molecule adsorbed on the local zigzag edges of an armchair-zigzag-edged GNR, a large vacancy-defected graphene, a ZGNR with an irregular edge, and a hydroxyl-functionalized ZGNR, respectively.

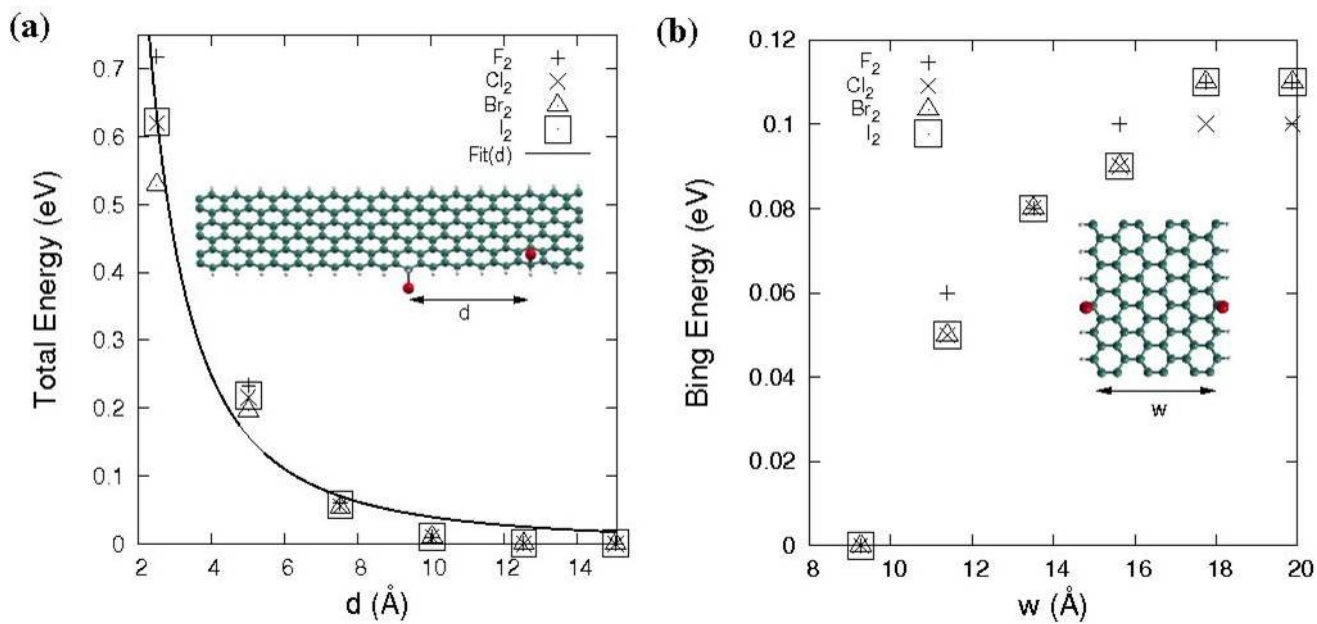

**Figure 2.** (a) The total energy of the optimized structures for the adsorption of two halogen atoms on the edge of the ZGNR as a function of the distance between the atoms where the energy of the lowest energy configuration is set to zero and the energy is fit by $a/d^2$ ($a=3.98$ eVÅ$^2$). The inset shows the optimized atomic geometries for two halogen atoms adsorbed on the edge of a ZGNR. (b) The calculated binding energy of a halogen molecule dissociatively adsorbed on the edges of the ZGNR as a function of the width where the smallest energy for each kind of the molecules is set to zero. The inset shows an optimized atomic structure for the adsorption of a halogen molecule on the edges of the ZGNR.

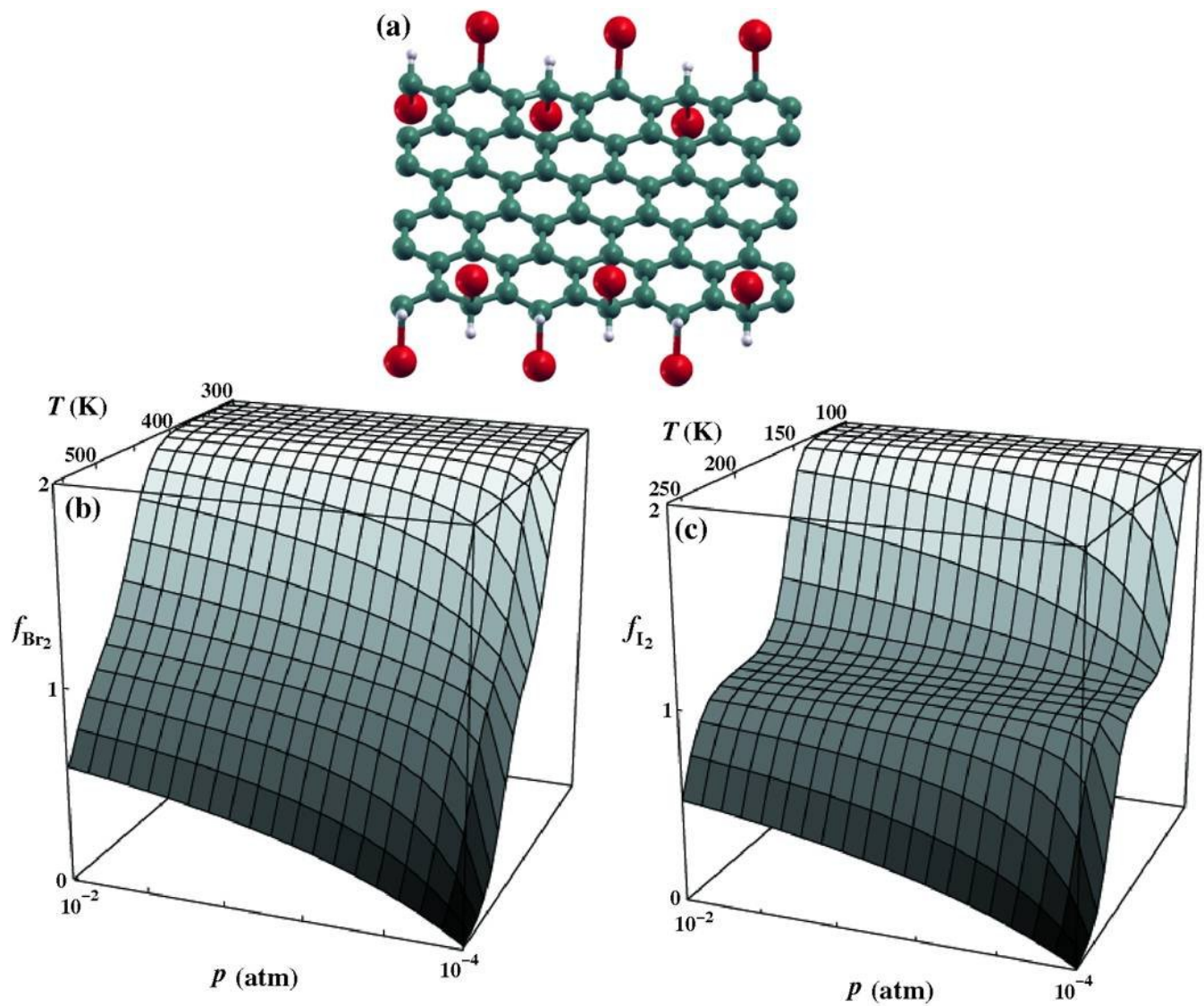

**Figure 3.** (a) Optimized atomic geometry for maximally adsorbed six halogen molecules on the edges of a ZGNR. (b,c) The occupation number of $Br_2$ or $I_2$ molecules on the edges of a ZGNR as a function temperature and the pressure.

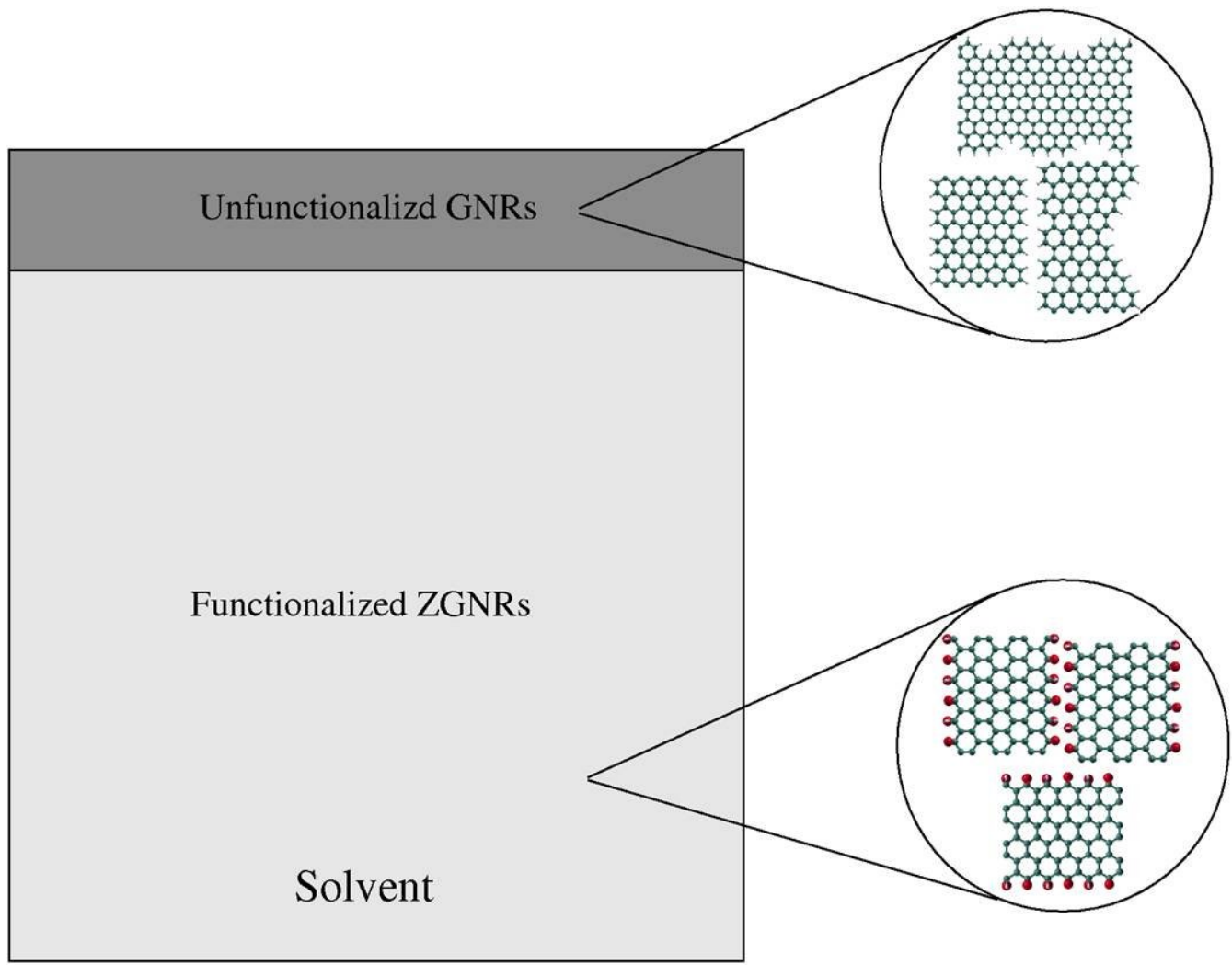

**Figure 4.** A schematic of the change of the solubility between functionalized ZGNRs and unfunctionalized GNRs for the separation of ZGNRs in a solvent where functionalized ZGNRs are soluble and unfunctionalized GNRs (e.g., AGNRs, ZGNRs with irregular edges, and armchair-zigzag-edged GNRs) are not soluble or vice versa.